\documentclass[twocolumn]{aastex631}
\usepackage{amsmath} 
\usepackage{graphicx}
\usepackage{booktabs}
\usepackage{color}
\usepackage{hyperref}

\newcommand{\Teff}{$T_{\mathrm{eff}}$}
\newcommand{\logg}{$\log g$}
\newcommand{\logH}{$\log \mathrm{H/He}$}
\newcommand{\chisq}{$\chi^{2}$}
\newcommand{\angstrom}{\text{\AA}}

\begin{document}

\title{White Dwarfs with Infrared Excess in the HET Dark Energy Experiment} 

\author[0009-0006-1574-5197]{Rudy A. Morales}
\affiliation{Department of Physics and Astronomy, Baylor University, Waco, TX 76798-7316, U.S.A.} 
\email{rudy\_morales1@baylor.edu}

\author[0000-0001-7010-7637]{Barbara G. Castanheira}
\affiliation{Department of Physics and Astronomy, Baylor University, Waco, TX 76798-7316, U.S.A.}

\author[0009-0003-9991-9438]{Jayden Blanchard}
\affiliation{Department of Physics and Astronomy, Baylor University, Waco, TX 76798-7316, U.S.A.}

\author[0000-0002-6164-6978]{Detlev Koester}
\affiliation{Institut f\"ur Theoretische Physik und Astrophysik, University of Kiel, 24098 Kiel, Germany}

\author[0000-0003-0963-0239]{P\'eter N\'emeth}
\affiliation{Astroserver.org, F\H{o} t\'er 1, 8533 Malomsok, Hungary}

\author[0000-0002-7470-5703]{S. O. Kepler}
\affiliation{Instituto de F\'{\i}sica, Universidade Federal do Rio Grande do Sul, 91501-970 Porto Alegre, Brazil}

\author[0000-0002-2307-0146]{Erin Mentuch Cooper}
\affiliation{Department of Astronomy, The University of Texas at Austin, 2515 Speedway Boulevard, Austin, TX 78712, USA}
\affiliation{McDonald Observatory, The University of Texas at Austin, 2515 Speedway Boulevard, Austin, TX 78712, USA}

\author[0000-0002-8433-8185]{Karl Gebhardt}
\affiliation{Department of Astronomy, The University of Texas at Austin, 2515 Speedway Boulevard, Stop C1400, Austin, TX 78712, USA}

\shorttitle{HETDEX White Dwarfs with IR Excess}
\shortauthors{Morales et al.}

\begin{abstract}

White dwarfs with infrared excess emission provide a window into the late stages of stellar evolution and the dynamics of circumstellar environments. Using data from the Hobby-Eberly Telescope Dark Energy Experiment (HETDEX), we characterized 30 white dwarfs exhibiting infrared excess, including 29 DA and 1 DB stars. While an infrared excess can arise from dusty disks or cool (sub-)stellar companions, our sample is limited to stellar companions due to our selection based on SDSS photometry, which is sensitive to excess emission at wavelengths $\lambda < 1\,\mu\mathrm{m}$. Our sample contains 22 newly identified excess sources not previously reported in the literature. Spectroscopic observations are available for 10 sources via SDSS, of which only 8 have prior spectroscopic classifications in the literature. 

In this paper, we present the determination of the effective temperature and surface gravity of these white dwarfs. We used the Balmer line profiles to compare with current atmospheric models to determine the photospheric parameters of the white dwarfs, minimizing contamination introduced by the infrared source. We used photometric data from the SDSS and the \textit{Gaia} mission to resolve the degeneracies between hot and cold solutions from spectroscopy, constraining the photospheric parameters. These results help refine our understanding of white dwarf evolution in binary systems, focusing on stellar companions that cause the infrared excess.

This study contributes to identifying systems with potential substellar companions or unresolved stellar partners, adding to the growing effort to map out the fate of planetary systems after their host stars evolve beyond the main sequence.
\end{abstract}


\section{Introduction} \label{sec:intro}

White dwarfs (WDs) represent the endpoint in stellar evolution for the vast majority of single stars \citep[e.g.][]{2008ARA&A..46..157W}, with a wide range of progenitor masses from 0.08\,$M_{\odot}$ to 8---10.5\,$M_{\odot}$ \cite[e.g.][]{2015ApJ...810...34W}. They provide a fundamental constraint for our understanding of stellar evolution and binary interaction. White dwarfs with infrared (IR) excess offer a unique opportunity to study the circumstellar environments of evolved stars and probe the final stages of planetary system evolution. 
These systems are also exciting targets for follow-up with facilities like JWST, which can spectroscopically characterize planetary debris and detect low-mass companions at mid-infrared wavelengths, as demonstrated by recent discoveries \citep[e.g.,][]{2024ApJ...962L..32M, 2025ApJ...981L...5F}. Our results add new candidates to the growing population of white dwarfs with infrared excess associated with unresolved stellar or substellar companions. Numerous studies have explored this phenomenon using data from WISE and other IR surveys \citep{2020ApJ...902..127X,2021ApJ...920..156L}, revealing that a significant fraction of WDs particularly those with low-mass companions—exhibit detectable excess emission at IR wavelengths.

When in binary systems --- both interacting and non-interacting --- white dwarfs may be the progenitors of a variety of interesting systems, including AM CVn systems, 
cataclysmic variables, and perhaps most interestingly, type Ia supernovae (SNe~Ia) \citep{1994ApJ...431..264I}. When the companion to the white dwarf evolves into a giant or supergiant, mass transfer through Roche lobe overflow can lead to nova outbursts or, if the white dwarf nears the Chandrasekhar limit, a thermonuclear explosion. Alternatively, the double-degenerate channel involves two white dwarfs and may also produce SNe~Ia \citep{2011ApJ...740L..18L}. These phenomena underscore the importance of white dwarfs for understanding stellar remnants, binary evolution, and explosive transients.

Infrared excess in white dwarfs is a powerful diagnostic of their surrounding environments and late-stage planetary system evolution. This excess emission can arise from one of three primary sources: a dusty circumstellar disk formed from tidally disrupted planetary bodies, a cool brown dwarf companion, or a low-mass main-sequence star such as an M dwarf \citep{2019MNRAS.489.3990R}. Recent studies suggest that dusty disks are detected in approximately 2–4\% of white dwarfs \citep{2014ApJ...786...77B, 2015MNRAS.449..574R,2019MNRAS.487..133W,2019MNRAS.489.3990R}, while brown dwarf companions account for about 0.5–2.0\% (\citealp{2021ApJ...920..156L} and references therein). M-dwarf companions occur in about 28\% of systems \citep{2011ApJ...729....4D}. Identifying the nature of the excess is crucial for understanding white dwarf binarity, planetary debris, and accretion processes.

A particularly well-studied case is G29-38, a pulsating white dwarf \citep{1974ATsir.810....1S} that also shows a significant infrared excess. Initially, \citet{1987Natur.330..138Z} proposed that the excess was due to a brown dwarf or planetary companion, but follow-up observations revealed a dusty disk composed primarily of amorphous olivine and a small amount of forsterite, located within 1–5\,$R_\odot$ of the white dwarf \citep{2005ApJ...635L.161R}. More recently, \citet{2022Natur.602..219C} used X-ray observations to infer an active accretion rate of planetary debris onto the white dwarf. This system exemplifies the kind of complex circumstellar environments that can be revealed through multiwavelength studies of infrared excess.

Using newly available spectra from HETDEX, in combination with complementary photometric and astrometric data from SDSS, \textit{Gaia}, 2MASS, and WISE, we analyzed a sample of 30 white dwarfs exhibiting infrared excess to determine their atmospheric parameters. For most of our targets, these HETDEX spectra represent their first spectroscopic measurements, providing a unique opportunity to characterize systems that have previously lacked detailed spectral analysis. In this paper, we describe the spectroscopic analysis of the HETDEX data, combined with all publicly available data from other surveys, such as the Sloan Digital Sky Survey \citep{2004ApJ...607..426K,2021MNRAS.507.4646K} and the Gaia mission \citep{2021MNRAS.508.3877G}. We discuss the fitting methods and the best determinations for the photospheric parameters of the white dwarfs.


\begin{figure}[t!] 
\centering 
\hspace*{-1.0cm} 
\includegraphics[scale=0.40]{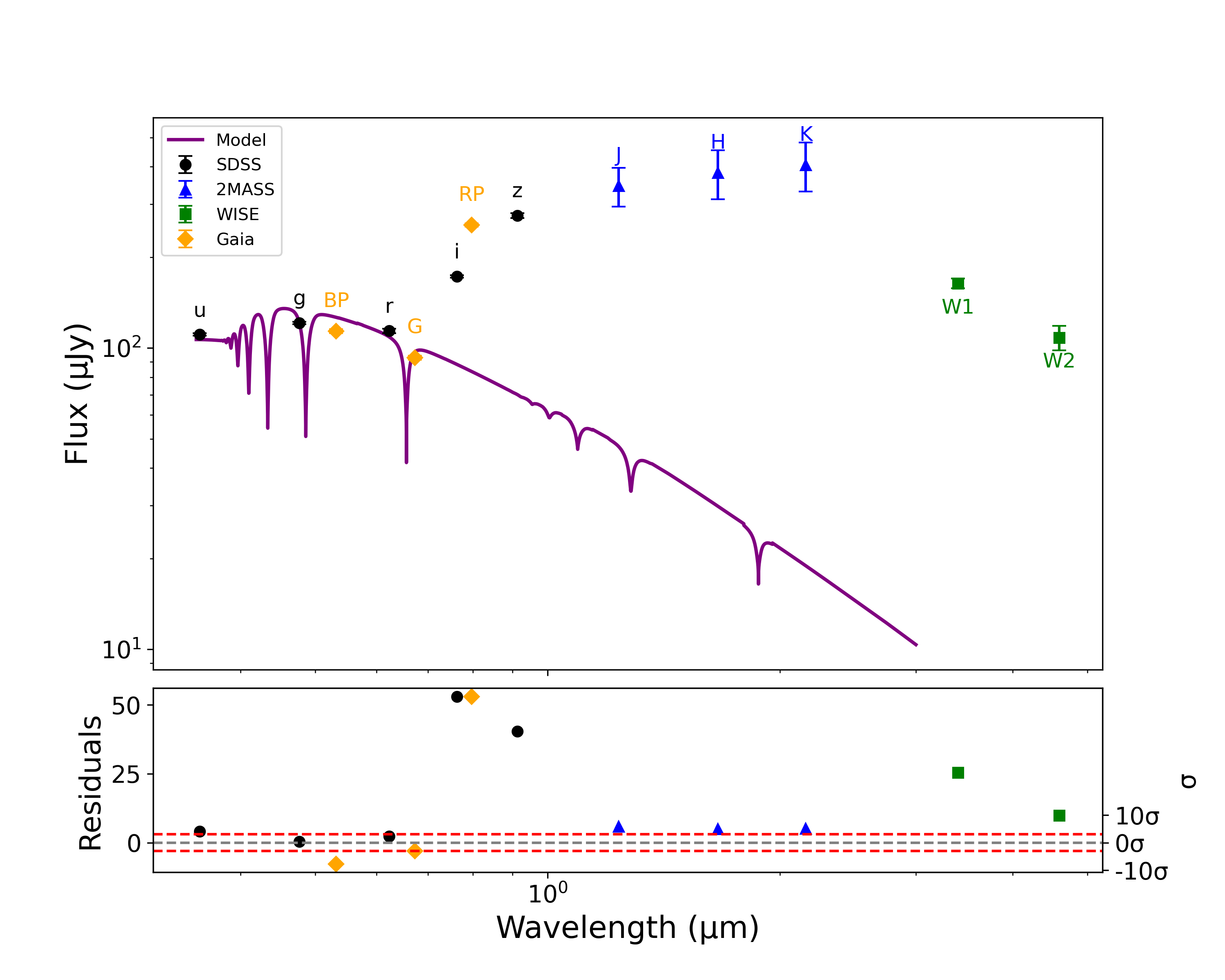} 
\caption{\textbf{Top Panel:} Spectral energy distribution (SED) fitting for the white dwarf SDSSJ113907.57+511103.6. The best-fit DA white dwarf model (purple) is compared to optical photometry from \textit{u, g, r, i, z} (black) and \textit{G, BP, RP} (orange diamonds). While the model closely matches the \textit{u, g, r} bands, significant deviations are present at \textit{i, z} and \textit{RP} (high outliers), as well as \textit{BP} and \textit{G} (low outliers). An IR excess is  detected  at \textit{J} (blue) just above 3$\sigma$ and \textit{W1, W2} (green) bands. \textbf{Bottom Panel:} Residuals expressed in units of measurement uncertainty ($\sigma$), with red dashed lines indicating the $\pm3\sigma$ threshold for identifying significant excess or deficit in the observed flux.}

\label{IR} 
\end{figure}


\section{Data}

\subsection{Sample Selection and HETDEX}

The primary goal of this study is to identify and characterize WDs with infrared (IR) excess signatures using spectroscopic and photometric data. This project is based on sources identified in the HETDEX survey \citep{2006NewAR..50..378H, 2011ApJS..192....5A,2021ApJ...923..217G}, which aims to probe dark energy by measuring the Hubble parameter, $H(z)$, and the angular diameter distance, $D_A(z)$, to explore the potential evolution of dark energy density \citep{2021ApJ...923..217G}. This is achieved by observing a large number of Lyman-$\alpha$ emitting galaxies with redshifts between $1.9 \leq z \leq 3.5$ across approximately $\sim 450$ deg$^2$ of the northern sky.
HETDEX utilizes the Visible Integral Field Replicable Unit Spectrograph (VIRUS; \citealt{2021AJ....162..298H}), composed of 156 low-resolution ($R \sim 750$) spectrographs covering the wavelength range 3500--5500 \AA, with a field of view of 22 arcmin, a segmented 11\,m primary mirror, and 35,000 fibers in 78 integral field units. Each fiber is separated by 2.5'', with a diameter of 1.5''. To ensure comprehensive flux sampling from all observing targets, the survey performs three dithered pointings of 6 minutes each.

Although HETDEX was not initially designed for stellar science, its magnitude-limited nature ensures that all continuum sources falling within the fibers are recorded. The Hobby–Eberly Telescope (HET) VIRUS Parallel Survey (HETVIPS) has reduced and made available to the community a continuum source catalog based on HETDEX observations \citep{2024ApJ...966...14Z}.

To identify white dwarfs within the HETDEX continuum catalog, we performed a positional cross-match with white dwarf catalogs from the Sloan Digital Sky Survey (SDSS; \citealt{2004ApJ...607..426K,2024yCat..75074646K}) and Gaia DR3 \citep{2021MNRAS.508.3877G}. The cross-match used sky coordinates (Right Ascension and Declination) with a maximum angular separation of 2 arcseconds to ensure accurate identification. This initial sample consisted of approximately 200 white dwarfs. No strict magnitude limits or quality cuts (e.g., photometric errors, parallax uncertainties, or astrometric flags) were applied at this stage; instead, the goal was to create a broad sample of WDs.   





\subsection{Identifying IR Excess White dwarfs}

\begin{figure}[b!]
    \centering
    \includegraphics[width=1\linewidth]{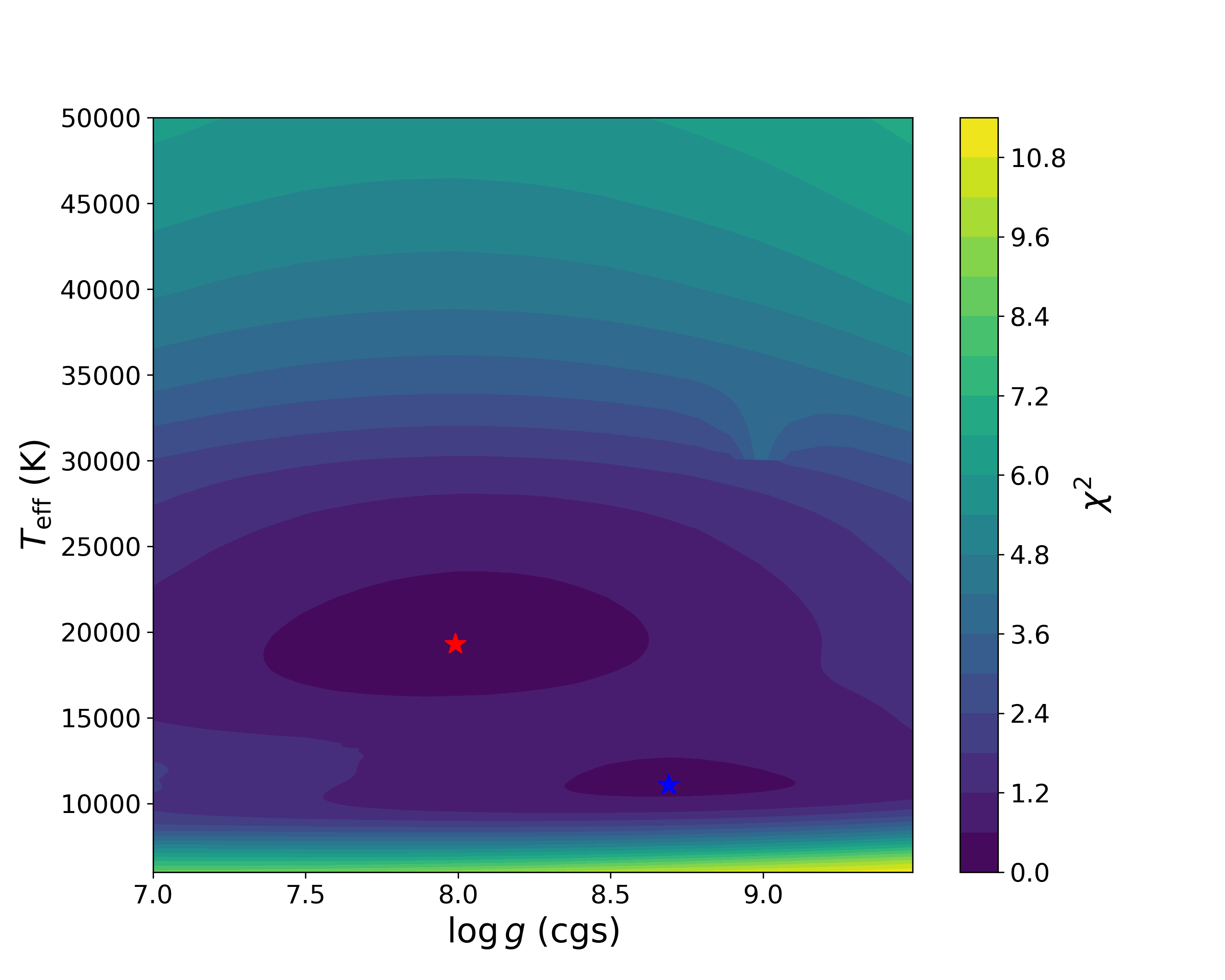}
    \caption{Contour plot of the $S$ values computed by comparing all the DA models in our grid to the observed spectrum of the star SDSS~J113907.57+511103.6. The red star represents the ``hot'' solution, while the blue star corresponds to the ``cold'' solution, reflecting the degeneracy in the Balmer lines fitting technique. }
  
    \label{fig:contourspectra}
\end{figure}

To confirm the presence of IR excess and ensure it was not an artifact of flux calibration or photometric uncertainties, we performed spectral energy distribution (SED) fitting using available optical and infrared photometry, following approaches in {e.g., \citet{2007ApJ...657.1013T} and \citet{2011MNRAS.417.1210G}}. Starting from an initial sample of approximately 200 white dwarfs identified as described above, we cross-matched these sources with the Two Micron All Sky Survey (2MASS; \citealt{2006AJ....131.1163S}) and the CatWISE catalog \citep{2020ApJS..247...69E,2021ApJS..253....8M}, which was constructed using data from the Wide-field Infrared Survey Explorer (WISE; \citealt{2010AJ....140.1868W}) and its extension NEOWISE \citep{2011ApJ...731...53M}, to obtain near- and mid-infrared fluxes. Cross-matching was performed using sky coordinates with a maximum angular separation of 2 arcseconds to ensure reliable associations.

To assess potential IR excess, we examined color-color diagrams (e.g., $u-g$, $g-r$, $r-i$, and $BP-RP$) and flagged sources that appeared redder than canonical white dwarf sequences. In addition, we considered SDSS targets flagged for red or IR excess, as these could indicate the presence of unresolved stellar or substellar companions, such as low-mass stars or brown dwarfs. Our analysis focuses on identifying such companions, as circumstellar dust or debris disks cannot be reliably detected in the optical bands ($\lambda < 1\,\mu\mathrm{m}$) used in our selection.

Table~\ref{tab:Photometry} summarizes the photometric properties of these objects, indicates which targets have been previously reported in the literature, and identifies those for which the first spectroscopic observation was obtained through HETDEX.

For each object, we constructed an SED using SDSS ($ugriz$), Gaia ($G$, $BP$, $RP$), 2MASS ($JHK_s$), and WISE ($W1$–$W2$) photometry where available. Each SED was fitted with a white dwarf atmospheric model anchored on the optical data. An IR excess was flagged only when at least one 2MASS or WISE $W1$ or $W2$ photometric point exceeded the model prediction by more than $3\sigma$, taking into account photometric uncertainties. The $W3$ and $W4$ bands were excluded from the IR excess selection criteria due to their broader point spread functions and lower sensitivity, which increase the likelihood of background contamination or source confusion \citep{2011ApJ...729....4D, 2020ApJ...891...97D}. This filtering reduced the sample to 30 white dwarfs with statistically significant IR excess. 

Figure~\ref{IR} illustrates this method using the example of SDSSJ113907.57+511103.6. The optical photometry from SDSS and Gaia generally aligns with the white dwarf model; however, elevated fluxes in SDSS $i$, $z$, and Gaia $RP$ suggest the possible presence of a low-mass companion, such as an M dwarf, contributing red excess. In the infrared, the 2MASS $J$ band lies just above the $3\sigma$ detection threshold, while WISE $W1$ and $W2$ show clear and significant excess relative to the model, consistent with an infrared excess. The bottom panel of Figure~\ref{IR} presents the residuals between the observed photometry and the white dwarf model, highlighting significant deviations at longer wavelengths consistent with the presence of IR-emitting sources, in our case being stellar companions.

Although our analysis excluded $W3$ and $W4$ from the selection criteria, it is important to interpret WISE-based excesses with caution. As shown by \citet{2020ApJ...891...97D}, IR excess detections from WISE—particularly for faint white dwarfs—can include false positives even after standard vetting procedures such as astrometric filtering and high-resolution imaging. No objects in our sample were flagged based on $W3$ or $W4$, as most sources are undetected or have only upper limits in these bands. We plan to obtain higher-resolution follow-up observations in the near- and mid-infrared (1.0–5.0 $\mu$m) using instruments such as the Gemini/GNIRS high-resolution IFU (R $\sim$ 1200–19000) to further investigate the nature of the IR excesses.

\begin{figure*}[t!]
    \centering
    \includegraphics[width=.85\linewidth]{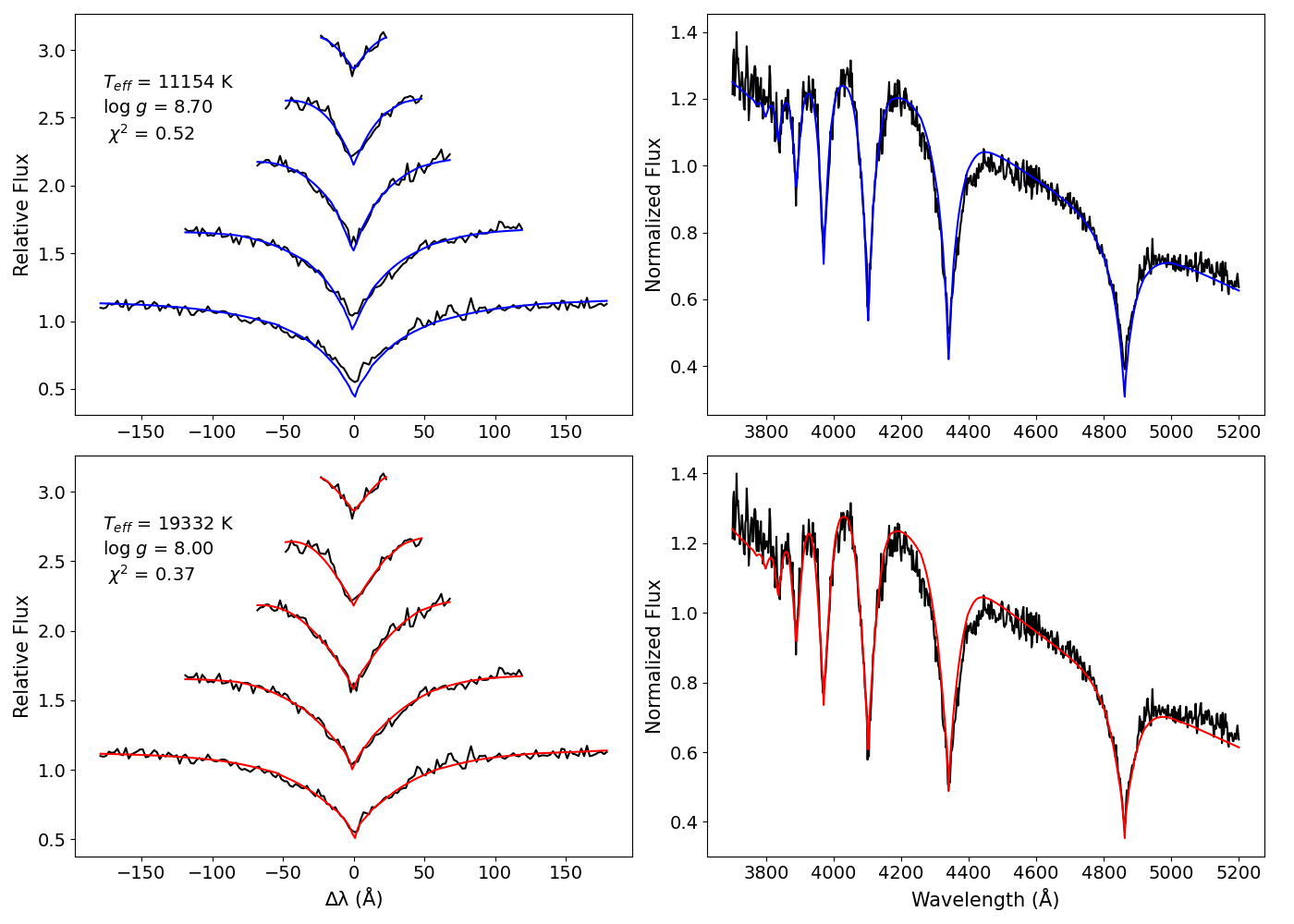}
    \caption{Example of our DA fitting procedure for SDSS~J113907.57+511103.6. 
    \textbf{Top Left Panel:} Best ``cold'' solution for Balmer line fitting, observed Balmer line profiles (black) and best model (blue) with \Teff\ = 11,154 K and \logg\ = 8.70. 
    \textbf{Top Right Panel:} Full observed spectrum with best cold model fitting, normalized at 4600\angstrom. 
    \textbf{Bottom Left Panel:} Best hot solution for Balmer line fitting, observed Balmer line profiles (black) and best model (red) with \Teff\ = 19,332 K and \logg\ = 8.00. 
    \textbf{Bottom Right Panel:} Full observed spectrum with best hot model fitting, normalized at 4600\angstrom.} 
    \label{fig:DA Fitting}
\end{figure*}

\begin{figure}[h!]
    \centering
    \includegraphics[width=1\linewidth]{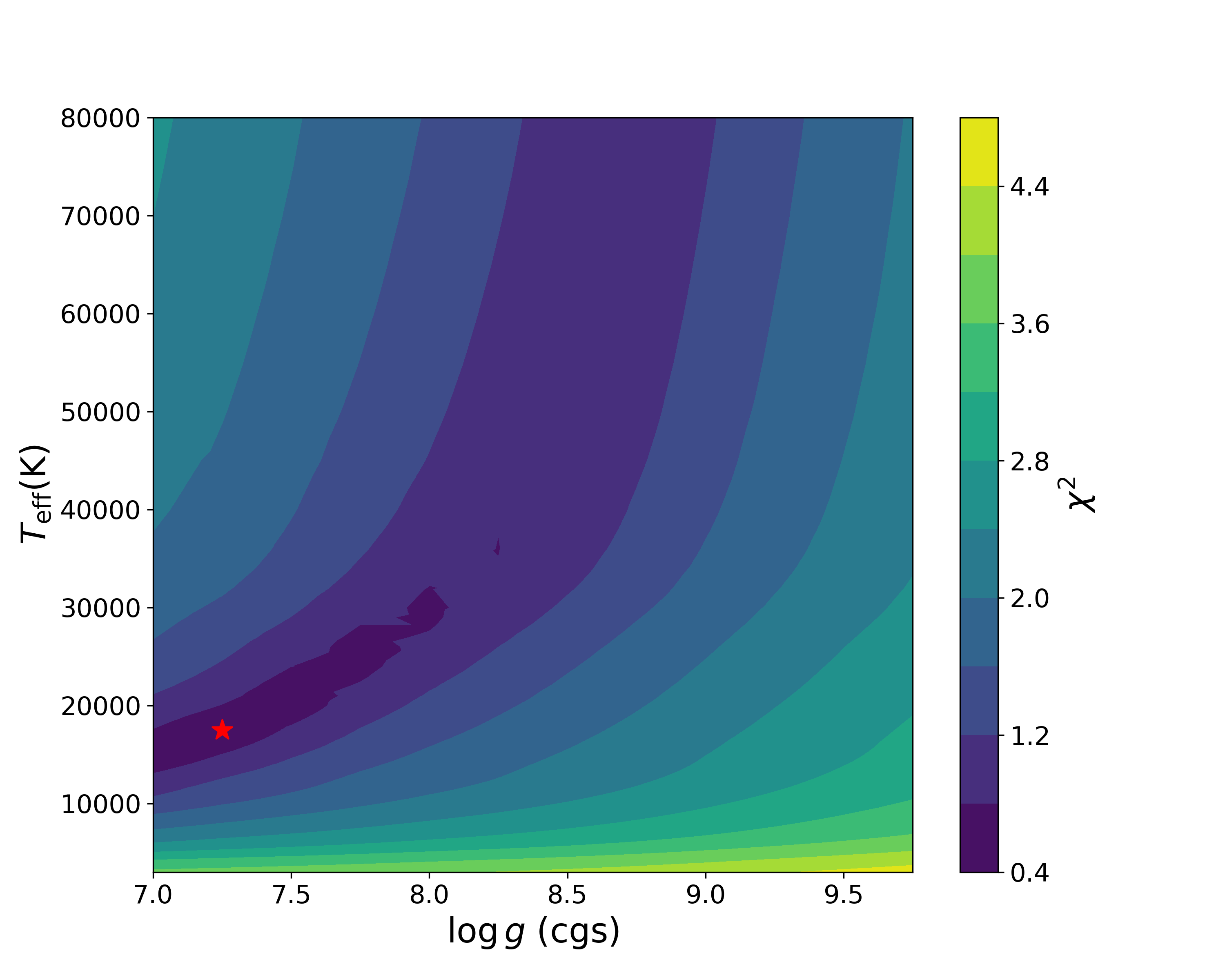}
    \caption{{Contour plot of the photometric fit using the \textit{u, g,} and \textit{r} bands for SDSS~J113907.57+511103.6. The fit is more sensitive to \Teff\ than \logg, with the best-matching model at \Teff = 17,193 K and \logg = 7.49 noted by the red star. The photometric constraints aided in selecting the preferred spectroscopic solution.}}
    \label{fig:phot}
\end{figure} 

\begin{table*}[t!]
\caption{SDSS names and photometry (u, g, r, i, z) and Gaia data (parallax and magnitudes in the bands G, BP, and RP) for the white dwarfs observed in the HETDEX survey with IR excess. Gaia and SDSS uncertainties are listed for each source. We verified that the majority of targets have Gaia Renormalized Unit Weight Error (RUWE), a metric assessing the goodness-of-fit of the Gaia astrometric solution, less than 1.4 and astrometric excess noise (AEN) near zero, indicating reliable astrometry. A subset of objects show elevated RUWE (up to 4.1) and AEN (up to 29 mas), likely due to unresolved companions or contamination. Reported parallaxes are adopted directly from Gaia DR3 and their uncertainties do not feature any contribution from excess noise.}
\label{tab:Photometry}
\centering
\hspace*{-2.6cm}
\resizebox{1.15\textwidth}{!}{
\begin{tabular}{c|ccccc|cccc|c}
\toprule\toprule
\textbf{Name SDSS J} & \textbf{u} & \textbf{g} & \textbf{r} & \textbf{i} & \textbf{z} & \textbf{$\Pi$} & \textbf{G} & \textbf{BP} & \textbf{RP} & \textbf{Notes}\\ 
& (mag) & (mag) & (mag) & (mag) & (mag) & (mas) & (mag) & (mag) & (mag)\\
\midrule
011302.50-002127.4 & 22.064$\pm$ 0.14 & 21.188$\pm$0.04 & 20.969$\pm$0.04 & 20.119$\pm$0.03 & 19.491$\pm$0.05 & 2.125$\pm$1.061 & 20.575$\pm$0.009 & 21.184$\pm$0.259 & 19.711$\pm$0.104 & (\(\ast\))\\ 
011322.56+015105 & 20.595$\pm$0.06 & 20.184$\pm$0.02 & 20.406$\pm$0.03 & 20.692$\pm$0.05 & 21.345$\pm$0.28 & 2.252$\pm$0.667 & 20.299$\pm$0.007 & 20.244$\pm$0.080 & 20.167$\pm$0.132 & (\(\ast\))\\ 
014009.00-001243.5 & 21.010$\pm$0.07 & 20.328$\pm$0.02 & 19.363$\pm$0.01 & 18.577$\pm$0.01 & 18.099$\pm$0.02 & 0.438$\pm$0.275 & 19.175$\pm$0.004 & 20.045$\pm$0.050 & 18.203$\pm$0.023 & (1),(3) \\  
014152.51-005241.6 & 20.344$\pm$0.05 & 19.773$\pm$0.01 & 19.263$\pm$0.01 & 18.297$\pm$0.01 & 17.659$\pm$0.02 & 1.365$\pm$0.216 & 18.793$\pm$0.003 & 19.724$\pm$0.067 & 17.798$\pm$0.027 & (1),(3) \\  
015225.39-005808.6 & 18.411$\pm$0.02 & 17.890$\pm$0.01 & 17.695$\pm$0.01 & 16.980$\pm$0.01 & 16.255$\pm$0.01 & 8.733$\pm$0.101 & 17.292$\pm$0.003 & 17.866$\pm$0.013 & 16.428$\pm$0.008 & (1),(5) \\ 
021309.19-005025.3 & 21.254$\pm$0.09 & 20.275$\pm$0.02 & 19.595$\pm$0.02 & 18.488$\pm$0.01 & 17.815$\pm$0.02 & 3.134$\pm$0.303 & 19.222$\pm$0.012 & 20.162$\pm$0.066 & 18.058$\pm$0.024 & (1),(3)\\ 
021855.87-013543.4 & 19.830$\pm$0.04 & 18.955$\pm$0.01 & 18.152$\pm$0.01 & 17.135$\pm$0.01 & 16.488$\pm$0.01 & 1.392$\pm$0.364 & 17.919$\pm$0.007 & 18.763$\pm$18.763 & 16.678$\pm$0.009 & (\(\ast\)) \\ 
021903.68-001733.9 & 19.383$\pm$0.02 & 19.250$\pm$0.01 & 18.920$\pm$0.01 & 18.107$\pm$0.01 & 17.502$\pm$0.01 & 1.513$\pm$0.193 & 18.516$\pm$0.004 & 19.227$\pm$0.039 & 17.644$\pm$0.019 & (1),(3) \\  
021938.57-001346.4 & 21.894$\pm$0.13 & 20.990$\pm$0.03 & 20.347$\pm$0.02 & 19.440$\pm$0.02 & 18.842$\pm$0.03 & 1.953$\pm$1.251 & 20.120$\pm$0.012 & 20.927$\pm$0.159 & 18.996$\pm$0.071 & (1),(2)\\ 
022125.36+001710.4 & 21.278$\pm$0.10 & 20.707$\pm$0.02 & 20.320$\pm$0.02 & 19.562$\pm$0.02 & 19.050$\pm$0.05 & -- & 20.364$\pm$0.021 & 20.7470.132 & 19.092$\pm$0.058 & (1),(3)\\  
110233.96+502739.9 & 21.734$\pm$0.12 & 20.933$\pm$0.03 & 20.314$\pm$0.02 & 19.264$\pm$0.02 & 18.551$\pm$0.03 & 0.478$\pm$0.411 & 19.826$\pm$0.005 & 20.673$\pm$0.100 & 18.695$\pm$0.029 & --\\ 
113907.57+511103.6 & 18.788$\pm$0.02 & 18.691$\pm$0.01 & 18.758$\pm$0.01 & 18.307$\pm$0.01 & 17.800$\pm$0.02 & 1.605$\pm$0.593 & 18.980$\pm$0.004 & 18.759$\pm$0.016 & 17.878$\pm$0.014 & (1),(4)\\ 
123354.89+521033.9 & 19.468$\pm$0.03 & 18.841$\pm$0.01 & 18.589$\pm$0.01 & 17.733$\pm$0.01 & 17.092$\pm$0.01 & 2.731$\pm$0.111 & 18.164$\pm$0.003 & 18.837$\pm$0.016 & 17.278$\pm$0.009 & (\(\ast\))\\ 
124158.58+544317.2 & 22.416$\pm$0.25 & 21.300$\pm$0.04 & 20.334$\pm$0.03 & 19.461$\pm$0.02 & 18.912$\pm$0.04 & 1.578$\pm$0.623 & 20.298$\pm$0.007 & 20.917$\pm$0.092 & 19.146$\pm$0.042 & (\(\ast\)) \\ 
125104.96+531727.2 & 21.875$\pm$0.19 & 21.027$\pm$0.04 & 20.458$\pm$0.03 & 19.475$\pm$0.02 & 18.897$\pm$0.04 & 1.011$\pm$0.387 & 20.045$\pm$0.005 & 20.860$\pm$0.120 & 19.068$\pm$0.036 & (\(\ast\)) \\ 
125351.57+510159.5 & 21.639$\pm$0.14 & 20.982$\pm$0.03 & 20.855$\pm$0.04 & 20.149$\pm$0.03 & 19.415$\pm$0.07 & -- & 20.733$\pm$0.021 & 21.060$\pm$0.153 & 19.517$\pm$0.160 & (\(\ast\)) \\ 
125656.08+540503.1 & 23.517$\pm$0.65 & 22.741$\pm$0.22 & 20.510$\pm$0.03 & 19.131$\pm$0.02 & 18.372$\pm$0.03 & 0.792$\pm$0.262 & 19.526$\pm$0.002 & 20.825$\pm$0.093 & 18.444$\pm$0.025 & (\(\ast\)) \\
125946.43+561320.2 & 20.235$\pm$0.04 & 19.579$\pm$0.01 & 19.327$\pm$0.01 & 18.517$\pm$0.01 & 17.737$\pm$0.02 & 4.648$\pm$0.179 & 18.856$\pm$0.003 & 19.505$\pm$0.045 & 17.853$\pm$0.022 & -- \\ 
130354.53+560844.0 & 22.208$\pm$0.21 & 21.444$\pm$0.05 & 21.287$\pm$0.07 & 20.635$\pm$0.05 & 20.008$\pm$0.10 & -- & 21.085$\pm$0.037 & 21.873$\pm$0.172 & 20.338$\pm$0.600 & (\(\ast\))\\ 
130418.62+511119.4 & 20.801$\pm$0.07 & 20.009$\pm$0.02 & 19.338$\pm$0.01 & 18.277$\pm$0.01 & 17.592$\pm$0.01 & 1.525$\pm$0.177 & 18.867$\pm$0.003 & 19.855$\pm$0.056 & 17.829$\pm$0.015 & (\(\ast\))\\ 
131315.85+535237.5 & 20.641$\pm$0.07 & 20.457$\pm$0.02 & 20.387$\pm$0.03 & 19.600$\pm$0.02 & 18.953$\pm$0.04 & -- & 20.440$\pm$0.015 & 20.377$\pm$0.086 & 19.177$\pm$0.060 & (\(\ast\))\\ 
133237.08+510213.1 & 20.320$\pm$0.04 & 19.477$\pm$0.01 & 18.693$\pm$0.01 & 17.875$\pm$0.01 & 17.395$\pm$0.01 & 0.767$\pm$0.210 & 18.606$\pm$0.005 & 19.240$\pm$0.027 & 17.492$\pm$0.010 & (\(\ast\)) \\ 
140035.96+540103.8 & 18.443$\pm$0.02 & 17.855$\pm$0.01 & 17.604$\pm$0.01 & 16.589$\pm$0.00 & 15.684$\pm$0.01 & 6.173$\pm$0.060 & 17.036$\pm$0.003 & 17.857$\pm$0.010 & 16.023$\pm$0.005 & (\(\ast\))\\ 
150959.39+511602.9 & 20.158$\pm$0.04 & 19.152$\pm$0.01 & 18.281$\pm$0.01 & 17.488$\pm$0.01 & 16.992$\pm$0.01 & 1.926$\pm$0.148 & 18.241$\pm$0.004 & 18.897$\pm$0.025 & 17.117$\pm$0.012 & (\(\ast\))\\ 
151523.32+504919.8 & 21.527$\pm$0.14 & 21.159$\pm$0.04 & 20.702$\pm$0.05 & 19.792$\pm$0.03 & 19.258$\pm$0.08 & 1.953$\pm$0.582 & 20.565$\pm$0.007 & 21.172$\pm$0.122 & 19.415$\pm$0.082 & (\(\ast\))\\ 
153343.45+535712.3 & 20.340$\pm$0.05 & 20.211$\pm$0.02 & 20.292$\pm$0.03 & 19.506$\pm$0.02 & 18.807$\pm$0.04 & 2.326$\pm$1.257 & 20.639$\pm$0.015 & 20.352$\pm$0.170 & 19.249$\pm$0.056 & (\(\ast\))\\ 
154540.48+515426.5 & 19.047$\pm$0.02 & 18.647$\pm$0.01 & 18.120$\pm$0.01 & 17.288$\pm$0.01 & 16.769$\pm$0.01 & 2.097$\pm$0.228 & 18.408$\pm$0.004 & 18.652$\pm$0.032 & 17.061$\pm$0.033 & (\(\ast\))\\ 
161051.80+503119.9 & 20.911$\pm$0.08 & 20.230$\pm$0.02 & 19.558$\pm$0.01 & 18.524$\pm$0.01 & 17.871$\pm$0.02 & 2.233$\pm$0.174 & 19.099$\pm$0.004 & 20.231$\pm$0.065 & 18.095$\pm$0.016 & (\(\ast\))\\ 
174517.73+670234.7 & 20.884$\pm$0.08 & 20.240$\pm$0.02 & 19.866$\pm$0.02 & 18.909$\pm$0.01 & 18.177$\pm$0.02 & 1.551$\pm$0.224 & 19.396$\pm$0.003 & 20.218$\pm$0.047 & 18.457$\pm$0.018 & (\(\ast\))\\ 
181204.26+650452.6 & 19.738$\pm$0.04 & 19.768$\pm$0.01 & 19.803$\pm$0.02 & 19.050$\pm$0.01 & 18.320$\pm$0.02 & 1.730$\pm$0.179 & 19.288$\pm$0.004 & 19.759$\pm$0.037 & 18.526$\pm$0.032 & (\(\ast\))\\
\bottomrule
\end{tabular}}
\vspace{0.4cm}
\begin{minipage}{\textwidth}
\small 
\textit{Notes:}  (\(\ast\)) Indicates targets for which the first spectroscopic observation was conducted by the HETDEX survey. (1) Targets are present in the SDSS-WDMS-Catalog \citep{2012MNRAS.419..806R}. Targets marked with following identifiers have prior spectroscopic classification in the literature, as compiled by the Montreal White Dwarf Database (MWDD); (2) 
\cite{2013ApJS..204....5K}. (3) \cite{2006AJ....132..676E, 2013ApJS..204....5K}. (4) \cite{2006AJ....132..676E,2011ApJ...730..128T,2013ApJS..204....5K}. (5) \cite{2006AJ....132..676E,2011ApJ...730..128T,2013ApJS..204....5K,2023MNRAS.521..760V}. 

\end{minipage}
\end{table*}

\section{Spectroscopic Analysis}

In this section, we describe our comparison of the HETDEX spectra for the white dwarfs identified in the continuum catalog that have infrared excess to atmospheric models. To accomplish the goals of HETDEX, the observations of Lyman-$\alpha$ emitting galaxies must be very accurately flux calibrated, which is not a trivial task in a telescope like the HET, with a fixed altitude, where the effective aperture of the mirror changes during the observations. \citet{2021ApJ...923..217G} discussed in detail their approach to model the throughput for HETDEX observations, testing their method by comparing the fluxes of various stars within their survey and SDSS spectra (see their Figure~16). Data from all sources observed by HETDEX have been systematically reduced by the HETDEX team \citep{2024ApJ...966...14Z}, which includes flux calibration of the continuum sources.

Although we benefit from spectra with well-calibrated flux, our initial fitting strategy focused exclusively on the normalized spectral lines rather than the full fluxed spectrum. This is consistent with our focus on probing stellar companions, which are the primary interpretation for the IR excesses in our sample. These secondary components can contribute significantly to the continuum level but do not typically affect the hydrogen or helium line profiles that are used to determine the white dwarf's photospheric parameters. Thus, by isolating and fitting only the hydrogen or helium lines, we ensure that the derived values of \Teff\ and \logg\ reflect the intrinsic properties of the white dwarf atmosphere.


\subsection{Models}

We have used an updated model grid based on the LTE atmospheric models described in \citet{2010MmSAI..81..921K} to determine the effective temperature (\Teff) and surface gravity (\logg) for white dwarfs with infrared excess. These models have been refined in subsequent works through improvements in the input physics, including updated Stark broadening profiles for hydrogen lines, improved treatment of convective energy transport using the ML2/$\alpha=0.8$ prescription (a mixing-length theory formulation where "ML2" refers to a specific choice of convective efficiency parameters and $\alpha$ is the mixing length to pressure scale height ratio), and more accurate opacities and equation-of-state treatments \citep[e.g.,][]{2019A&A...628A.102K, 2020A&A...635A.103K}.


The grid of DA models spans $6,000\,\mathrm{K} \leq T_{\mathrm{eff}} \leq 50,000\,\mathrm{K}$ and $7 \leq \log g \leq 9.5$ (cgs units), and was calculated for ML2/$\alpha=0.8$. The grid of DB models is slightly different, ranging from $3,000\,\mathrm{K} \leq T_{\mathrm{eff}} \leq 70,000\,\mathrm{K}$, $7 \leq \log g \leq 9.5$ (cgs units), and relative abundances of hydrogen $-9 \leq \log \mathrm{H/He} \leq 0$. In our sample, all stars have spectral type DA, except for one DB (discussed in Section~\ref{dbfit}).

\begin{figure}[b!]
    \centering
    \includegraphics[width=1\linewidth]{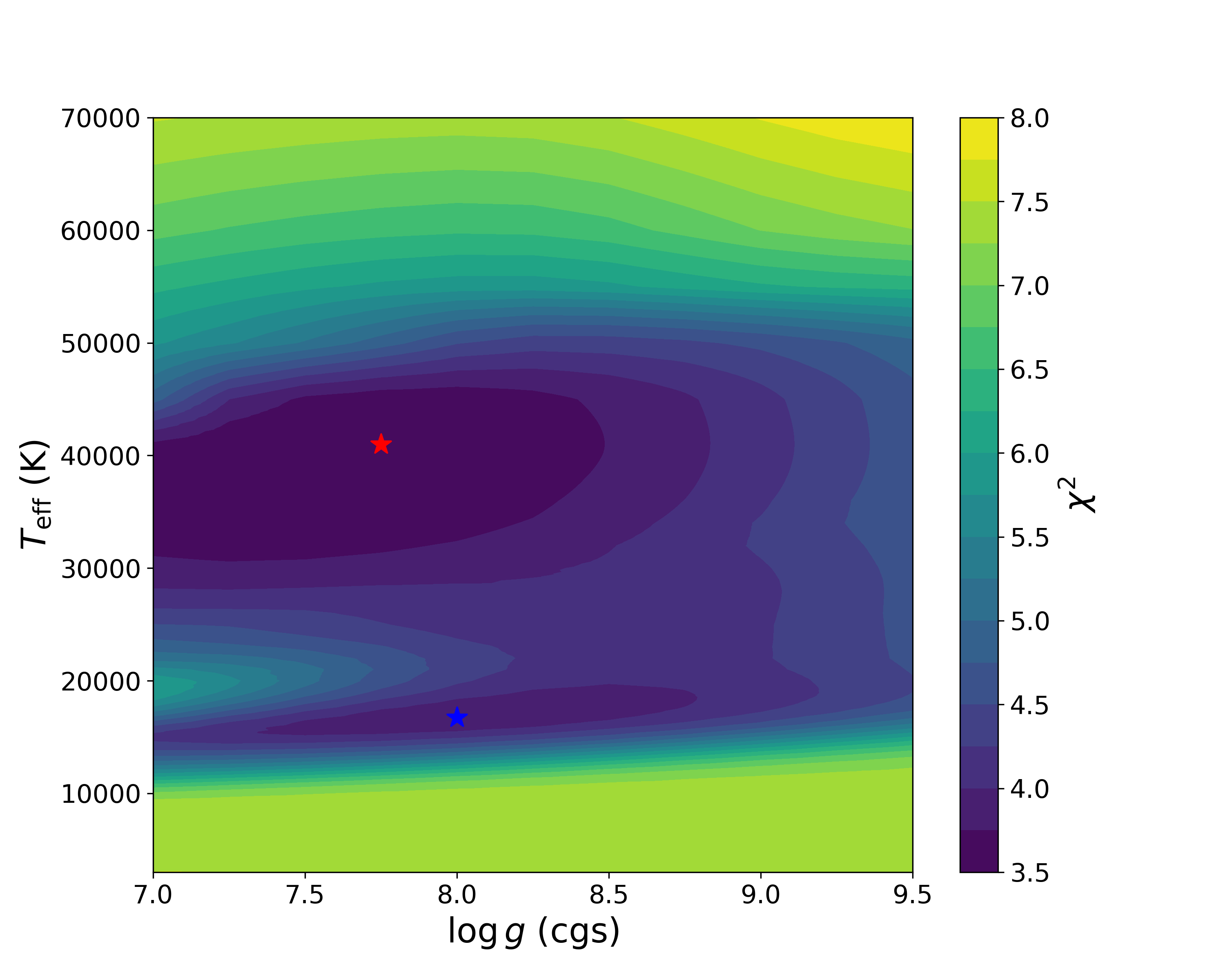}
    \caption{Contour plot of the $S$ values computed by comparing all the pure DB models in our grid to the observed spectrum of the star SDSS~J181204.26+650452.6, showing a similar degeneracy, with the red and blue stars marking the ``hot'' and ``cold'' solutions, respectively.}
    \label{fig:contourspectradb}
\end{figure}

\subsection{Fitting DA White Dwarfs}

HETDEX observations cover a fixed wavelength range that includes the Balmer series from H$\beta$ to H$\eta$. While H$\eta$ is technically within this range, we exclude it from our fitting due to the combination of reduced line strength at higher-order transitions and increased sensitivity \citep{2006MNRAS.372.1799K} to theoretical uncertainties in line broadening at these wavelengths. This approach follows established spectroscopic techniques that limit fitting to H$\beta$ through H$\zeta$ to ensure robust and consistent parameter determination \citep{1992ApJ...394..228B,2005ApJS..156...47L,2011ApJ...743..138G,2011ApJ...730..128T,2013A&A...559A.104T}.

Our spectroscopic analysis begins by centering each Balmer line to ensure the isolation of the line profiles. Proper centering is essential to minimize systematic errors that affect the subsequent fitting process. Once centered, the spectral regions containing these Balmer lines were isolated to avoid introducing artifacts that could compromise the normalization process \citep{1992ApJ...394..228B,1995ApJ...449..258B}. The central wavelengths and total fitting ranges used for each Balmer line are listed in Table~\ref{tab:balmerlines}.

Following isolation, we then normalize the continuum around each line by fitting a linear function to the line wings and dividing the observed flux by this continuum fit. This procedure sets the continuum level to unity, allowing the line profile to be analyzed independently of any residual flux calibration or broadband shape in the spectrum. The lines in the DA models were isolated and normalized in the same way as the spectroscopic data for each star, ensuring that model and observed line profiles could be compared on a consistent, continuum-normalized scale for accurate parameter fitting. This normalization procedure differs from the one used for DB white dwarfs, where the continuum is first defined by fitting a synthetic model spectrum multiplied by a high-order polynomial across a broader spectral region.

\begin{table}[h!]
\centering 
\caption{Center of the rest wavelength ($\lambda$) and extent of the wings ($\lambda$ Range) of the Balmer lines used in our spectroscopic fits for DA white dwarfs. All quantities are in \AA.}
\label{tab:balmerlines}
\hspace{-1cm} 
\begin{tabular}{lcc}
\toprule\toprule
{Balmer Line} & {Rest Wavelength ($\lambda$)} & {$\lambda$ Range}\\ 
\midrule
   H$\beta$ & 4861.350\,\angstrom & $\pm 180\,\angstrom$ \\ 
   H$\gamma$ & 4340.472\,\angstrom & $\pm 120\,\angstrom$ \\ 
   H$\delta$  &  4101.734\,\angstrom & $\pm 60\,\angstrom$ \\ 
   H$\epsilon$ &  3970.075\,\angstrom & $\pm 40\,\angstrom$ \\ 
   H$\zeta$ &  3889.069\,\angstrom & $\pm 30\,\angstrom$  \\ 
   H$\eta$ &   3835.397\,\angstrom & $\pm 20\,\angstrom$ \\ 
   \bottomrule
\end{tabular}
\end{table}

Additional processing steps before line fitting include interpolation of the model spectral fluxes onto the observed wavelength grid and convolution with a Gaussian profile to replicate instrumental resolution ($R \sim 750$) and observational conditions. These steps ensure that the spectral models are consistent with the observational conditions, thereby enabling a more precise fitting of the models to the data. 

We fit the spectroscopic lines using a least-squares approach, as described in \citet{1986ApJ...305..740Z}, by minimizing the squared differences ($S$) between the observed spectra ($I_\mathrm{obs}$) and the model spectra ($I_\mathrm{calc}$):

\begin{equation}
S = \sum_{i=1}^n \left(I_{\mathrm{obs}} - I_{\mathrm{calc}}\right)_i^2
\label{eq:chi_squared}
\end{equation}

\noindent
In this formulation, we assume uniform weighting (i.e., $W^{-2} = 1$) across all data points. The minimum value of $S$ is denoted as $S_0$. Following \citet{1986ApJ...305..740Z}, we estimate the uncertainties ($\sigma$) in \Teff\ and \logg\ using nearby points in the model grid.

\begin{equation}
\sigma^2=\frac{d^2}{S-S_0}
\label{eq:sigma}
\end{equation}

\begin{table*}[t!]
\caption{Spectroscopy and photometry fitting parameters for the sample. The \textbf{Spectroscopy} columns include both the hot and cold solutions for \Teff~and \logg~(and \logH~for the DB star). The \textbf{Photometry} columns list the best-fit parameters from the photometric data. The \textbf{final} adopted \Teff~and \logg~values, determined from the photometric analysis shown in bold.}
\label{tab:Parameters}
\small
\centering
\hspace*{-2.3cm} 
\begin{tabular}{lcccccccc}
\toprule\toprule
\textbf{Name SDSS J} & \multicolumn{5}{c}{\textbf{Spectroscopy}} & \multicolumn{2}{c}{\textbf{Photometry}} \\ 
\cmidrule(lr){2-6} \cmidrule(lr){7-8}
& \textbf{\Teff (Hot)}K & \textbf{\logg (Hot)} & \textbf{\Teff (Cold)}K & \textbf{\logg (Cold)} & \textbf{\logH (Hot, Cold)} & \textbf{\Teff}K & \textbf{\logg} \\ 
\midrule
011302.50-002127.4 & 16,657 (546) & 6.95 (0.05) & \textbf{9,841 (143)} & \textbf{7.22 (0.023)} & --- & 6,466 & 7.05 \\ 
011322.56+015105.4 & 15,236 (237) & 7.89 (0.136) & \textbf{13,335 (406)} & \textbf{7.78 (0.101)} & --- & 12,172 & 8.11 \\
014009.00-001243.5 & \textbf{25,233 (1328)} & \textbf{7.69 (0.167)} & 9,781 (198) & 8.79 (0.173) & --- & 24,953 & 7.10 \\ 
014152.51-005241.6 & 19,123 (694) & 8.18 (0.152) & \textbf{14,430 (759)} & \textbf{8.55 (0.136)} & --- & 11,411 & 7.23 \\
015225.39-005808.6 & 29,081 (855) & 7.13 (0.168) & \textbf{8,833 (130)} & \textbf{8.49 (0.164)} & --- & 7,008 & 7.35 \\ 
021309.19-005025.3 & 20,599 (820) & 7.09 (0.157) & \textbf{9,629 (143)} & \textbf{7.98 (0.201)} & --- & 6,314 & 7.06 \\
021855.87-013543.4 & 26,834 (1,136) & 7.10 (0.155) & \textbf{8,885 (125)} & \textbf{8.19 (0.189)} & --- & 15,063 & 7.02 \\ 
021903.68-001733.9 & 30,930 (782) & 7.97 (0.256) & \textbf{9,148 (171)} & \textbf{9.40 (0.133)} & --- & 12,020 & 7.89 \\ 
021938.57-001346.4 & 18,277 (655) & 8.08 (0.123) & \textbf{13,673 (687)} & \textbf{8.50 (0.103)} & --- & 6,695 & 7.03 \\
022125.36+001710.4\((\ast)\) & 19,501 (751) & 7.38 (0.172) & \textbf{9,963 (150)} & \textbf{8.15 (0.143)} & --- & 6,468 & 8.00 \\ 
110233.96+502739.9 & 29,829 (752) & 7.58 (0.013) & \textbf{8,910 (140)} & \textbf{8.80 (0.010)} & --- & 17,345 & 7.00 \\ 
113907.57+511103.6 & \textbf{19,332 (729)} & \textbf{8.00 (0.136)} & 11,154 (354) & 8.70 (0.180) & --- & 17,193 & 7.49 \\ 
123354.89+521033.9 & 16,253 (624) & 8.10 (0.116) & \textbf{15,018 (432)} & \textbf{8.02 (0.116)} & --- & 9,966 & 7.31 \\
124158.58+544317.2 & 23,979 (923) & 7.97 (0.154) & \textbf{9,894 (148)} & \textbf{8.77 (0.096)} & --- & 6,923 & 7.08 \\ 
125104.96+531727.2 & 20,779 (827) & 7.01 (0.090) & \textbf{9,704 (138)} & \textbf{7.84 (0.146)} & --- & 9,281 & 7.30 \\
125351.57+510159.5\((\ast)\) & 15,053 (368)  & 8.48 (0.115) & \textbf{13,023 (617)} & \textbf{8.42 (0.106)} & --- & 6,797 & 8.00 \\ 
125656.08+540503.1 & 19,166 (735) & 7.12 (0.129) & \textbf{9,795 (147)} & \textbf{7.85 (0.141)} & --- & 7,836 & 7.07 \\
125946.43+561320.2 & 32,833 (927) & 7.26 (0.215) & \textbf{8,264 (128)} & \textbf{8.78 (0.136)} & --- & 6,465 & 7.33 \\ 
130354.53+560844.0\((\ast)\) & 15,100 (316) & 7.21 (0.121) & \textbf{14,122 (641)} & \textbf{7.14 (0.123)} & --- & 6,237 & 8.00 \\ 
130418.62+511119.4 & 21,166 (852) & 8.08 (0.137) & \textbf{14,706 (705)} & \textbf{8.88 (0.133)} & --- & 9,738 & 7.27 \\
131315.85+535237.5\((\ast)\) & 20,012 (755) & 7.76 (0.136) & \textbf{10,617 (224)} & \textbf{8.60 (0.134)} & --- & 8,549 & 8.00 \\ 
133237.08+510213.1 & \textbf{25,015 (1,150)} & \textbf{7.73 (0.183)} & 9,618 (168) & 8.67 (0.177) & --- & 20,997 & 7.01 \\
140035.96+540103.8 & 19,556 (727) & 7.88 (0.295) & \textbf{10,983 (293)} & \textbf{8.55 (0.170)} & --- & 8,216 & 7.34 \\ 
150959.39+511602.9 & 22,345 (972) & 7.11 (0.237) & \textbf{9,377 (144)} & \textbf{8.08 (0.136)} & --- & 15,063 & 7.04 \\
151523.32+504919.8 & 28,141 (983) & 7.60 (0.289) & \textbf{9,223 (147)} & \textbf{8.76 (0.605)} & --- & 7,075 & 8.12 \\ 
153343.45+535712.3 & 21,573 (897) & 8.20 (0.142) & \textbf{11,045 (371)} & \textbf{9.10 (0.130)} & --- & 15,215 & 8.52 \\
154540.48+515426.5 & 21,240 (861) & 7.60 (0.269) & \textbf{10,463 (222)} & \textbf{8.65 (0.129)} & --- & 12,705 & 7.43 \\ 
161051.80+503119.9 & 26,682 (1284) & 7.41 (0.162) & \textbf{9,436 (152)} & \textbf{8.68 (0.200)} & --- & 7,532 & 7.28 \\ 
174517.73+670234.7 & 17,745 (615) & 7.60 (0.146) & \textbf{15,200 (200)} & \textbf{7.32 (0.118)} & --- & 9,586 & 7.19 \\
181204.26+650452.6\((\dagger)\) & 40,961 (607) & 7.73 (0.052) & \textbf{16,828 (162)} & \textbf{8.02 (0.148)} & -2.16 (0.052), \textbf{-6.39 (0.190)} & 15,332 & 7.22 \\ 
\bottomrule
\end{tabular}
\begin{minipage}{\textwidth}
\small
\noindent\textit{Notes:} Targets marked with an asterisk (\(\ast\)) lack parallax measurements. For these objects, the photometric fitting was performed using the color index method, where \Teff~was estimated by comparing the observed \(u - g\) and \(u - r\) colors to Koester model predictions, assuming a fixed \logg~= 8.0. (\(\dagger\)) indicates the only DB white dwarf in our sample.
\end{minipage}
\end{table*}

Here, $S_0$ corresponds to the best-fit solution and $S$ represents the next-best minimum. Following the approach of \citet{1986ApJ...305..740Z}, we choose the parameter $d$ such that the difference in \Teff\ and \logg\ is within approximately 5\% of the best-fit values. This threshold balances the need to avoid unrealistically small denominators in the error propagation when adjacent grid points have nearly identical \chisq\ values while still reflecting meaningful variation in the model parameters.

In Figure~\ref{fig:contourspectra}, we show the contour plot of the comparison between all DA models in our grid with the observed spectra for the star SDSS~J113907.57+511103.6. The red and blue stars indicate the two minima, which we refer to as ``hot'' and ``cold'' solutions. 

This is a well-known degeneracy in Balmer line fitting, where more than one combination of effective temperature and surface gravity can produce similar Balmer line profiles. This occurs because the Balmer lines first deepen and broaden with increasing temperature, reaching maximum strength near 13,000–15,000 K due to optimal hydrogen excitation \citep{2007MNRAS.382.1377R, 2019ApJ...871..169G, 2019ApJ...882..106G}, but then weaken at higher temperatures as hydrogen becomes increasingly ionized and fewer atoms remain at the n = 2 level to produce Balmer absorption. A similar degeneracy is observed in DB white dwarfs, where neutral helium lines peak in strength around 25,000 K \citep{2015A&A...583A..86K, 2019ApJ...871..169G} and weaken at both higher and lower temperatures due to changes in ionization balance. As a result, both ``hot'' (e.g., ~20,000 K) and ``cold'' (e.g., ~10,000 K) models can produce comparably good fits, particularly when data quality or wavelength coverage is limited.

In Figure~\ref{fig:DA Fitting}, we show the comparison between the observed Balmer line profiles (left panels) and the total spectrum (right panels) for the star SDSS~J113907.57+511103.6 and the best models for the ``cold'' solution (top panels) and the ``hot'' solution (bottom panels). The ``cold'' solution is a for model with \Teff$=11,154$\,K and \logg$=8.70$\,dex, while the ``hot'' solution model has \Teff$=19,332$\,K and \logg$=8.00$\,dex. Line profiles have been normalized as described in this section, while the total spectrum has been normalized at 4600\angstrom. 

The goodness of fit for the solutions ``hot'' and ``cold'' are nearly indistinguishable by visual inspection and very similar in \chisq; therefore, we resort to an external determination to select our best solution.

We restricted our analysis to the u, g, and r bands due to contamination from the companion in the i and z bands. In principle, we could use the spectral energy distribution based on photometric measurements to distinguish between different temperatures. However, since our targets were selected for their infrared excess, the companion significantly contaminates the red and infrared filters, making them unreliable for this analysis \citep{2006AJ....131..571H}. 

To adjust for the transformation between the observed SDSS magnitudes and the AB magnitude system, we applied a correction to the \textit{u} band using \( u = u_{\text{SDSS}} - 0.040 \) from \citep{2006AJ....132..676E}.

When Gaia parallax is available, we estimate distances by inverting the parallax and convert the observed apparent magnitudes in the \textit{u, g,} and \textit{r} bands to absolute magnitudes using the standard distance modulus equation. These absolute magnitudes are then compared with the Koester photometric models calculated for this work. In cases where no parallax is available, we instead use the color index method, comparing the observed \( u - g \) and \( u - r \) colors to photometric models while assuming \( \log g = 8.0 \). This approach follows the same methodology as for spectroscopy, applying equation~\ref{eq:chi_squared} to fit and computing uncertainties according to equation~\ref{eq:sigma}. 

The assumption of $\log g = 8.0$ is motivated by the fact that the mass distribution of white dwarfs strongly peaks near $0.6~M_{\odot}$, which corresponds to $\log g \approx 8.0$ for typical white dwarf radii \citep[e.g.,][]{2007MNRAS.375.1315K}. This provides a reasonable approximation of the photometric parameters in the absence of independent distance estimates, without introducing significant bias for the effected stars in the sample.

In Figure~\ref{fig:phot}, we present the contour plot that compares the observed photometry of SDSS~J113907.57+511103.6 to all models in our grid, based on fitting only the \textit{u, g,} and \textit{r} bands after converting to absolute magnitudes using Gaia parallax. Since this approach relies on only three photometric data points from the spectral energy distribution, it is significantly more sensitive to \Teff, than to \logg, which remains poorly constrained. The best matching model to the photometry and parallax in our grid corresponds to \Teff = 17,193 K and \logg = 7.49. 

It is important to emphasize that our photometric determinations serve as an external constraint to distinguish between the ``hot'' and ``cold'' solutions derived from Balmer line fitting. For SDSS~J113907.57+511103.6, this method supports the ``hot'' spectroscopic solution as the best fit, resulting in \Teff = 19,332 K and \logg = 8.00.

We have carried out a similar analysis for all our DAs in our sample, and the results are listed in Table~\ref{tab:Parameters}. A detailed composite-spectrum analysis of SDSS~J113907.57+511103.6 using the {\sc XTgrid} fitting code is presented in Appendix~\ref{app:xtgrid}, which supports the hot solution and provides additional constraints on the companion.

\begin{figure*}[t!]
    \centering
    \includegraphics[width=.85\linewidth]{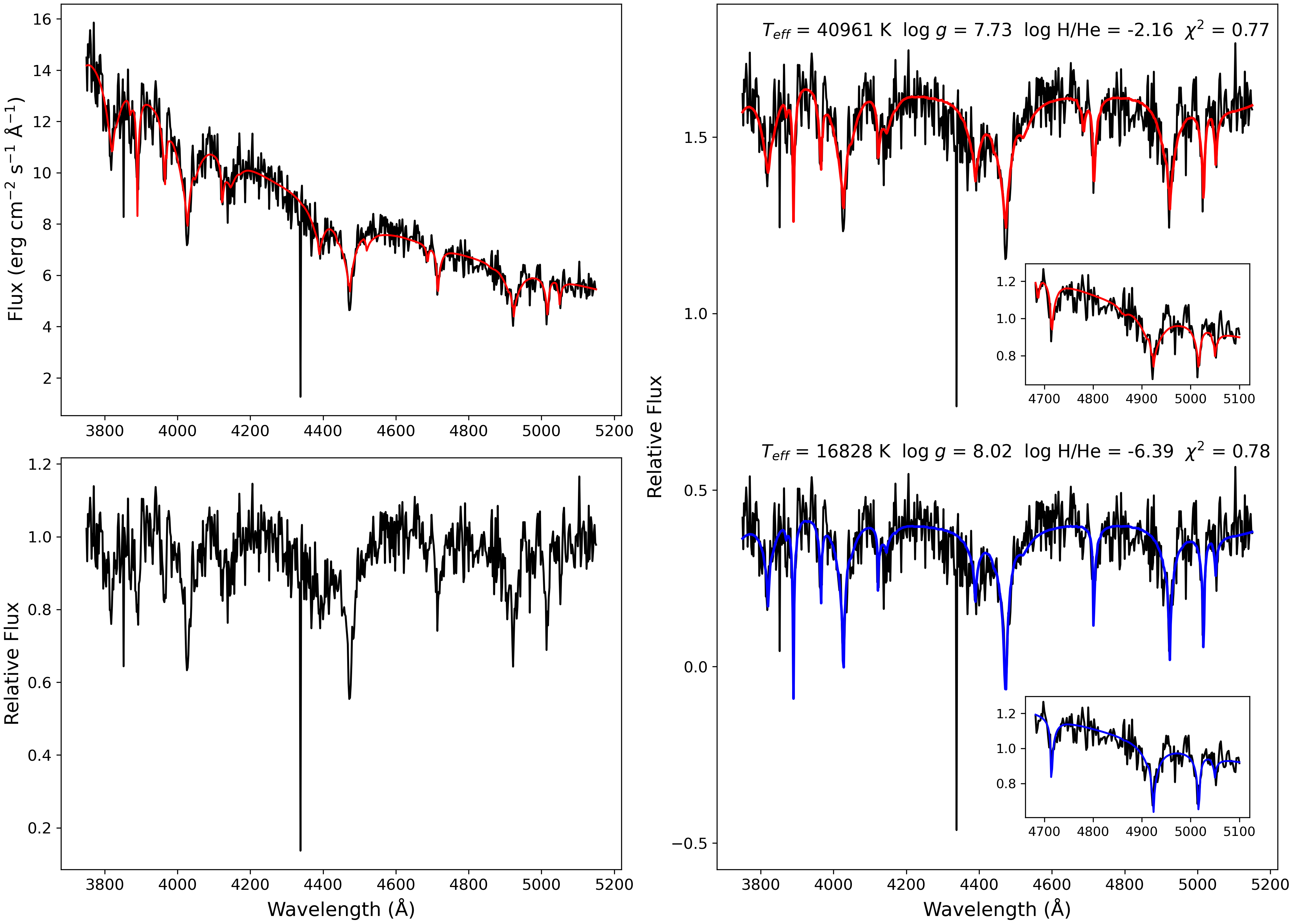}
    \caption{Example of our DB white dwarf fitting procedure for SDSS~J181204.26+650452.6, the only DB white dwarf in our dataset. \textbf{Top Left Panel:} The observed spectrum (black) with the flux calibration function is applied to determine the continuum. The red line represents the scaled model used for this calibration step. \textbf{Bottom Left Panel:} The normalized spectrum after dividing by the continuum function, showing the absorption features in relative flux units. \textbf{Right Panel:} The best-fit models overlaid on the observed spectrum, with the ``hot'' solution (\Teff\ = 40961 K, \logg\ = 7.73, \logH\ = -2.16) shown in red and the ``cold''  (\Teff\ = 16828 K, \logg\ = 8.02, \logH\ = -6.39)  in blue. The models are offset for clarity, and the inset highlights the H$\beta$ region (4681–5100 \angstrom) used for determining the hydrogen abundance.}
    \label{fig:DB Fitting}
\end{figure*}

\subsection{Fitting DB White Dwarfs}\label{dbfit}

For the single DB white dwarf in our dataset, we followed a spectroscopic technique similar to that described by \citet{2011ApJ...737...28B}, which differs from the fitting process used for DA white dwarfs by an additional parameter, namely the hydrogen abundance \citep{2019ApJ...882..106G}. As DB white dwarfs have helium-dominated photospheres, the spectral analysis focuses primarily on the neutral helium absorption lines, but if present, hydrogen lines must also be considered.

We begin the fitting process by normalizing the flux of the observed spectra using our DB model grid to define the continuum. Each observed spectrum is initially fitted with a synthetic model spectrum multiplied by a high-order polynomial (typically fifth order) to account for residual flux calibration effects. The best fitting combination of model and polynomial is determined by minimizing the chi-squared between the observed and model fluxes in the wavelength range 3750---5150\AA. This polynomial scaled model provides a smoothed representation of the continuum, from which we extract flux values at predefined anchor points (e.g., 3750, 3900, 4210, 4610, 4810 and 5110,\AA). We then linearly interpolate between these anchor points to define the continuum, which is used to normalize the observed spectrum by dividing by this function, effectively setting the continuum to unity.

After normalization, the observed spectrum is fitted with a pure helium DB model grid to determine \Teff\ and \logg, again focusing on the blue portion of the spectrum where neutral helium lines provide the strongest constraints \citep{2019ApJ...871..169G}. As in our DA fitting procedure, all model spectra are first convolved with a Gaussian kernel to match the instrumental resolution of the observations ($R \sim 750$) prior to comparison. 

In Figure~\ref{fig:contourspectradb}, we show the contour plot of the $S$ values computed by comparing all pure DB models in our grid to the observed spectrum of the star SDSS~J181204.26+650452.6. As with the DAs, the DB shows a similar degeneracy, which are indicated by the red and blue star for the ``hot'' and ``cold'' solutions, respectively.


The \logH\ abundance was determined by fixing the \Teff\ and \logg\ parameters while varying \logH. This procedure was performed iteratively: after an initial estimate of the hydrogen abundance, we refit the spectrum to refine \Teff\ and \logg. We repeated this process until convergence was achieved for all three parameters, accounting for the covariance between them \citep{2011ApJ...737...28B,2015A&A...583A..86K,2019ApJ...871..169G}.

Our estimated H$\beta$ region ($4681 - 5100$\,\angstrom) was used instead of the typical H$\alpha$ region ($6400 - 6800$\,\angstrom) due to the spectral coverage in our sample \citep{2018ApJ...857...56R,2019ApJ...871..169G}.

To lift the degeneracy, we also used the photometric information. The results of our analysis for the only DB in our sample is in Table~\ref{tab:Parameters}.

\section{Conclusion and Future Prospects}\label{conclusion}

\begin{figure}[t!]
    \centering
    \hspace*{-0.9cm}
    \includegraphics[width=1.2\linewidth]{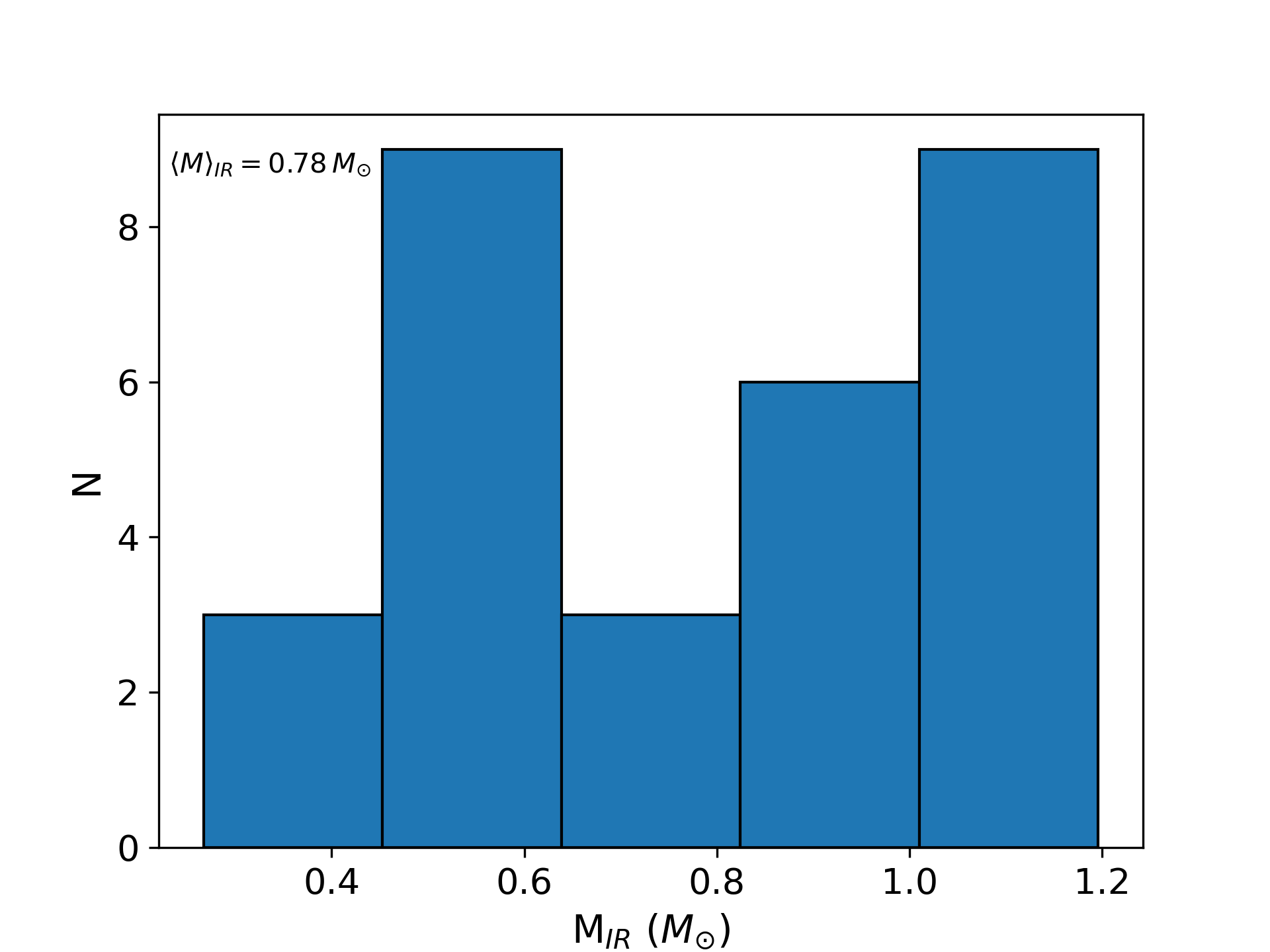}
    \caption{Mass distribution histogram for the infrared-excess DA white dwarfs in our sample. The distribution shows a mean mass of $\langle M\rangle_{\mathrm{IR}}=0.78\,M_{\odot}$, indicating a tendency towards higher mass white dwarfs compared to the local DA population within 40\,pc.}
    \label{fig:mass}
\end{figure}

We have presented a spectroscopic analysis of 30 white dwarfs exhibiting infrared excess, identified within the Hobby-Eberly Telescope Dark Energy Experiment continuum catalog. The presence of IR excess in these objects, typically arising from unresolved cool stellar or sub-stellar companions, can introduce contamination that complicates spectral fitting. Using only the absorption lines in the observed spectra and the photometric measurements, we could determine \Teff\ and \logg\ for these objects. In addition to the photospheric parameters, we confirm spectroscopically that these objects are white dwarfs. 
Our objects comprise a valuable list for follow-up ground-based observations to fully characterize the source of IR excess. Moreover, white dwarfs with stellar companions provide a unique opportunity to probe the effects of stellar evolution on close binary systems and the fate of stellar companions after the main sequence. These systems are ideal targets for JWST and future telescopes to resolve the IR excess or detect stellar companions.

A direct comparison of the atmospheric parameters derived from our infrared excess white dwarfs in Table~\ref{tab:Parameters} with DA white dwarfs in the 40 pc sample \citep{2024MNRAS.527.8687O} complied from the Montreal White Dwarf Database \citep{2017ASPC..509....3D} highlights several interesting trends. From our analysis, we find mean values of $\langle$\Teff$\rangle_{\mathrm{IR}}=12464$\,K, $\langle$\logg$\rangle_{\mathrm{IR}}=8.26$, and $\langle M \rangle_{\mathrm{IR}}=0.78\,M_{\odot}$. In contrast, the corresponding means for the local (40\,pc) DA population are $\langle$\Teff$\rangle_{\mathrm{40pc}}=8650$\,K, $\langle$\logg$\rangle_{\mathrm{40pc}}=8.08$, and $\langle M \rangle_{\mathrm{40pc}}=0.65\,M_{\odot}$. This indicates that the infrared-excess sample, on average, has significantly higher effective temperatures and surface gravities thus larger masses compared to the broader nearby DA white dwarf sample, as shown in Figure~\ref{fig:mass}. These differences might suggest younger cooling ages or higher-mass progenitor systems for the IR-excess white dwarfs, possibly indicating more recent or active planetary system interactions. Additionally, we note that 9 of our targets have masses above $1\,M_{\odot}$ (excluding the single DB in our sample). However, these high masses should be interpreted cautiously since, as discussed in \cite{2019MNRAS.489.3990R}, such objects could potentially be non-DA white dwarfs, leading to significant uncertainties in their derived atmospheric parameters and masses.

\section{Acknowledgments}
\label{acknowledgments}

NSF research grant number 2108737, Exploring the galaxy and its white dwarfs with the HET dark energy experiment. 

HETDEX (including the WFU of the HET) is led by the University of Texas at Austin McDonald Observatory and Department of Astronomy with participation from the Ludwig-Maximilians- Universität München, Max-Planck-Institut für Extraterrestriche-Physik (MPE), Leibniz-Institut für Astrophysik Potsdam (AIP), Texas A\&M University, Pennsylvania State University, Institut für Astrophysik Göttingen, The University of Oxford, Max-Planck-Institut für Astrophysik (MPA), The University of Tokyo, and Missouri University of Science and Technology. In addition to Institutional support, HETDEX is funded by the National Science Foundation (grant AST- 0926815), the State of Texas, the US Air Force (AFRL FA9451-04-2- 0355), and generous support from private individuals and foundations.

Based on observations obtained with the Hobby-Eberly Telescope (HET), which is a joint project of the University of Texas at Austin, the Pennsylvania State University, Ludwig-Maximilians-Universitaet Muenchen, and Georg-August Universitaet Goettingen. The HET is named in honor of its principal benefactors, William P. Hobby and Robert E. Eberly.

VIRUS is a joint project of the University of Texas at Austin, Leibniz-Institut fur Astrophysik Potsdam (AIP), Texas A\&M University (TAMU), Max-Planck-Institut fur Extraterrestrische Physik (MPE), Ludwig-Maximilians-Universitaet Muenchen, Pennsylvania State University, Institut fur Astrophysik Goettingen, University of Oxford, and the Max-Planck-Institut fur Astrophysik (MPA). In addition to Institutional support, VIRUS was partially funded by the National Science Foundation, the State of Texas, and generous support from private individuals and foundations. 

We acknowledge the Texas Advanced Computing Center (TACC) at the University of Texas at Austin for providing computing resources that have contributed to the research results reported in this paper. URL: http://www.tacc.utexas.edu/. The authors thank Patrick Dufour for a fruitful discussion on the line profile fitting.

This publication makes use of data products from the Wide-field Infrared Survey Explorer, which is a joint project of the University of California, Los Angeles, and the Jet Propulsion Laboratory/California Institute of Technology, and NEOWISE, which is a project of the Jet Propulsion Laboratory/California Institute of Technology. WISE and NEOWISE are funded by the National Aeronautics and Space Administration.

P.N. acknowledges support from the Grant Agency of the Czech Republic (GA\v{C}R 22-34467S).
This research has used the services of \mbox{\url{www.Astroserver.org}} under reference GRTO5U.

\appendix
\renewcommand{\thefigure}{\Alph{section}\arabic{figure}}
\setcounter{figure}{0}

\section{Disentangling with {\sc XTgrid}}
\label{app:xtgrid}

This appendix presents an independent spectral analysis of the composite-spectrum binary in the sample, SDSS~J113907.57+511103.6. 
We include it here to demonstrate the viability of such an approach for future studies of composite systems.

We applied the steepest descent data-driven spectral analysis procedure {\sc XTgrid} \citep{2012MNRAS.427.2180N} to evaluate the validity of the hot and cold solutions for SDSS~J113907.57+511103.6. 
The procedure constructs a theoretical WD+MS composite spectrum and iteratively adjusts the surface parameters of both binary components along with their flux contributions to the final spectrum. 
All parameters are refined to simultaneously match the detailed optical spectrum and the spectral energy distribution (SED) of the system. 
Various constraints can be incorporated to help break the strong degeneracies inherent in modeling composite spectra.

A key challenge in this analysis was the uncertainty in the distance to SDSS~J113907.57+511103.6, as well as the unknown mass ratio and spectral types of the binary components. 
To address this, we assumed the mass of the DA white dwarf to be 0.6\,M$_\odot$, consistent with the peak of the DA mass distribution, and adopted its radius from the mass-radius relation. 
That radius allowed us to estimate the relative flux contributions to the composite spectrum, enabling a systematic search for the best-matching effective temperature (\Teff), surface gravity (\logg), and metallicity of the companion. 
We also assumed the system to be a bound physical binary. 
The adopted WD mass was then used to determine the system's distance, from which the absolute magnitude, luminosity, and radius of both components were derived. 
{\sc XTgrid} iteratively applies these constraints in a $\chi^2$ minimization procedure until the global minimum is reached.

\begin{figure*}[]
    \centering
    \includegraphics[width=1\linewidth]{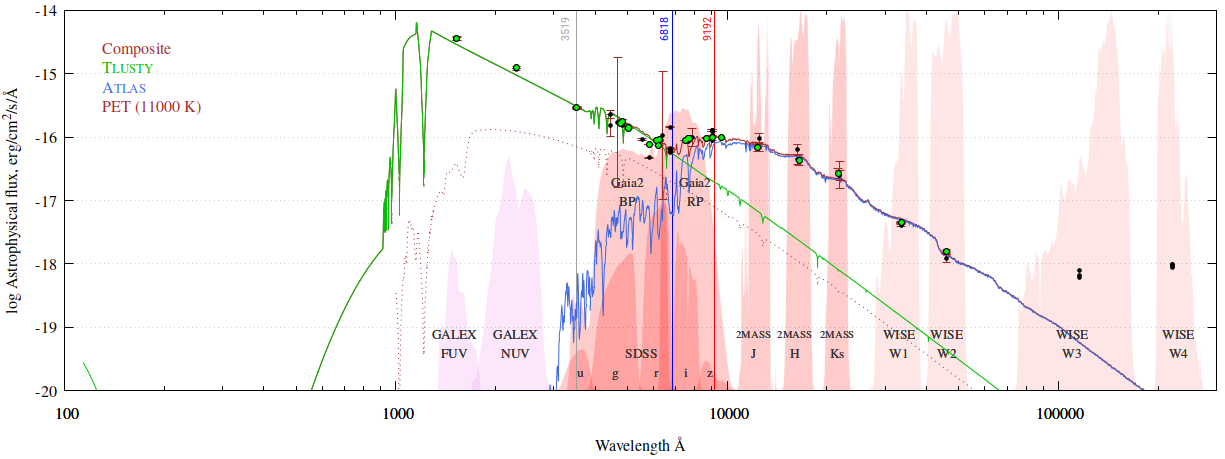}
    \caption{
     Spectral energy distribution (SED) of SDSS~J113907.57+511103.6, showing photometric observations (black and green points with error bars) overlaid with the best-fitting composite spectral model (brown). 
     The individual contributions from the white dwarf ({\sc Tlusty} model, green line) and the M-dwarf companion ({\sc Atlas} model, blue line) are also shown. 
     Filter transmission curves for GALEX, SDSS, Gaia, 2MASS, and WISE bands are shaded in the background.
     The WISE W3 and W4 band fluxes are affected by zodiacal foreground emission. 
     The model successfully reproduces both the ultraviolet and optical flux, as well as the infrared excess, confirming the presence of a cool main-sequence companion. 
     The theoretical composite SED was normalized to the observed data in the SDSS $u$-band at 3519\,\AA. For comparison, an 11,000 K WD model is shown (PET, brown dashed line) from the synthetic spectral library of \cite{2013A&A...559A.104T}. 
     }
    \label{fig:xtgrid}
\end{figure*}

The composite model consists of a DA spectrum calculated with the {\sc Tlusty} non-Local Thermodynamic Equilibrium (non-LTE) stellar atmosphere code \citep{1995ApJ...439..875H, 2017arXiv170601859H}, and a synthetic spectrum of an M-type main-sequence star extracted from the BOSZ \citep{2024A&A...688A.197M} spectral library calculated with the MARCS LTE code \citep{2008A&A...486..951G}.

We complemented the HETDEX spectrum with an optical spectrum from the Sloan Digital Sky Survey (SDSS DR12) \citep{2015ApJS..219...12A} that covers the 3800-9000 \AA\ spectral range, and allows a good sampling of both binary members. 
We ran a fit with the initial constraints and started with a 19,000\,K DA WD model with a fixed surface gravity ($\log g = 8.0$).
We found that a 3000\,K M-dwarf provides a self-consistent and nearly perfect match to both the SED and the available spectra next to an 18,350\,K DA WD.
Next, we repeated the analysis using the previous results as starting values and let all parameters change freely. 
This resulted in $T_{\rm eff}=18170\pm550$ K, and $\log{g}=7.80\pm0.16$ for the WD, as shown in Figure~\ref{fig:xtgrid}. 
However,  systematic errors likely supersede the statistical errors.

In contrast, repeating the procedure with an 11,000\,K DA model with $\log g = 8.7$ could not reproduce the SED. 
Such a cool, compact, and low-luminosity WD would require a substellar companion, which is inconsistent with the observed infrared excess of the system.
Even without a spectral decomposition, a cool DA WD cannot reproduce the UV slope of the SED as indicated by the GALEX FUV/NUV and SDSS $u$-band fluxes in Figure~\ref{fig:xtgrid}.

With the currently available observational data, our approach was sufficient to distinguish between the previously identified hot and cool solutions. 
A more comprehensive analysis will require time-resolved spectroscopy and knowledge of the dynamic mass ratio.


\bibliography{sample631}{}
\bibliographystyle{aasjournal}

\end{document}